\documentclass[fleqn,twoside]{article}
\usepackage{espcrc2}
\usepackage{amssymb}
\usepackage{psfig}
\usepackage{epsfig}

\usepackage{graphicx}
\usepackage[figuresright]{rotating}

\title{Deep Inelastic Neutrino Interactions\thanks{Work 
under Contract Nos.~DE-AC02-98CH10886 and FG02-91ER40664 
with the U.S.~Department of Energy.
}}

\author{S. Kretzer\address{Physics Department, Brookhaven National Laboratory,\\
Upton, New York 11973, U.S.A.}\address{
RIKEN-BNL Research Center, Bldg. 510a, Brookhaven 
National Laboratory,\\
Upton, New York 11973 -- 5000, U.S.A.
}\
and M.H. Reno\address{Department of Physics and Astronomy, University of Iowa,\\
Iowa City, Iowa 52242 USA
}}

\begin{document}

\begin{abstract}
Understanding neutrino interactions is an important task in 
searches for neutrino oscillations; 
e.g.~the $\nu_{\mu} \to \nu_{\tau}$ oscillation hypothesis
will be tested through $\nu_\tau$ production of $\tau$ 
in long-baseline experiments as well as
underground neutrino telescopes.
An anomaly in the deep inelastic interaction of neutrinos 
has recently been observed by the NuTeV collaboration
-- resulting in a measured weak mixing angle $\sin^2 \Theta_{\rm W}$
that differs by $\sim 3 \sigma$ from the standard 
model expectation.
In this contribution to the proceedings of NUINT02, we
summarize results on the NLO neutrino
structure functions and cross sections in which charm quark mass
and target mass effects in the collinear approximation are
included.
\vspace{1pc}
\end{abstract}

\maketitle

\section{INTRODUCTION}
\label{sec:intro}

The interpretation of the
atmospheric muon neutrino deficit in the SuperKamiokande
is that muon neutrinos oscillate to tau neutrinos
with nearly maximal mixing \cite{Fukuda:1998ub}. 
Experiments such as ICANOE, OPERA and MONOLITH will be able to
test the oscillation hypothesis by searching for tau neutrino
conversion to taus \cite{lble} after the accelerator beam
of muon neutrinos has passed over the long-baseline between
CERN and the Gran Sasso Laboratory. The MINOS experiment run
at lower energies \cite{minos} will look for the characteristic muon neutrino
deficit. Ultimately, one would like to know the neutrino-nucleon
charged current cross section to the level of a few percent
for precision measurements of oscillation mixing angles.
Mass corrections, from both the out-going charged lepton and from
the target nucleon, are important at this level. Here, we report
on lepton and target mass corrections in the collinear limit
to neutrino-nucleon charged current interactions in deep inelastic
scattering. Next-to-leading order QCD corrections  and charm
production are included.
This is a portion of the total cross section which
includes quasi-elastic scattering and few pion production at
the energies of interest 
\cite{py,by,gallagher}. 

At the precision level, an anomaly in the interaction of neutrinos of higher 
${\cal{O}}(100\ {\rm GeV})$ energies has recently
been observed by the FERMILAB NuTeV experiment \cite{mcf}. For this
Paschos-Wolfenstein-like \cite{pw} measurement of the weak mixing angle 
$\sin^2 \Theta_{\rm W}$, deep inelastic scattering is the
dominant interaction channel. NLO QCD, charm mass ($m_c$) and target mass
corrections are potentially important ingredients in the
interpretation of the experimental result.

In the next section, we review the deep inelastic scattering (DIS)
formalism,
concentrating on charged current scattering (CC) where two additional
structure functions $F_4$ and $F_5$ come into play for energies
not too large compared to the outgoing lepton mass. We show our
results for the structure functions with NLO QCD corrections, and
$m_c$ and nucleon mass $M_N$ corrections in the collinear limit, in
Section 3. Our cross section results for $\nu_\tau$ CC scattering
with isoscalar nucleons $N$ and $\nu_\mu\, N$ CC scattering are shown
in Section 4. Conclusions appear in Section 5. More complete
references
and further details of this work appear in Ref. \cite{kretzerreno},
and extensions to include neutral current interactions as well as
target mass corrections through the operator
product expansion approach of Georgi and Politzer \cite{gp}, through NLO QCD,
will be presented in Ref. \cite{soon}.

\section{DIS FORMALISM}
\label{sec:formulae}     

The QCD dynamical part of the evaluation of neutrino interactions 
is given through the hadronic tensor
\begin{eqnarray} \label{eq:wmunu0} 
W_{\mu\nu} & \equiv&
{1\over {2 \pi}}\ \int e^{iq\cdot z}d^4 z 
\langle N|J_\mu (z) J_\nu (0)
|N\rangle
\\ \nonumber
%\label{eq:wmunu2}
& =& {1\over {\pi}}\ {\rm Disc} \int e^{iq\cdot z}d^4 z 
\langle N|iT \left(J_\mu (z) J_\nu (0)\right)
|N\rangle \\ \nonumber
%\label{eq:wmunu3}
&=& -g_{\mu \nu}W_1 + {p_\mu p_\nu\over M^2}W_2 %\\ \nonumber
%&&
-i\epsilon_{\mu\nu\rho\sigma}
{p^{\rho}q^\sigma\over M^2}W_3 \\ \nonumber
&
+ &{q_\mu q_\nu \over M^2}W_4+{p_\mu q_\nu+p_\nu q_\mu\over M^2} W_5\ .
\end{eqnarray}

Sandwiched between the current operators in the first line of 
Eq.~\ref{eq:wmunu0}, 
an inclusive sum over a complete set of states has been performed:
\begin{equation}
{\bf 1}=\sum_X \mid X \rangle \langle X \mid
\end{equation}
in which 
\begin{eqnarray*}
\mid X\rangle & =& \mid N\rangle\quad {\rm for \ quasielastic}\\ 
\mid X\rangle & =& \mid N \, n\pi \rangle\quad{\rm
resonances/few\ particles}\\
\mid X\rangle & = &\mid {\rm multiparticle}\rangle \quad {\rm DIS\ continuum}\ .
\end{eqnarray*} 
\begin{figure}[t]
\psfig{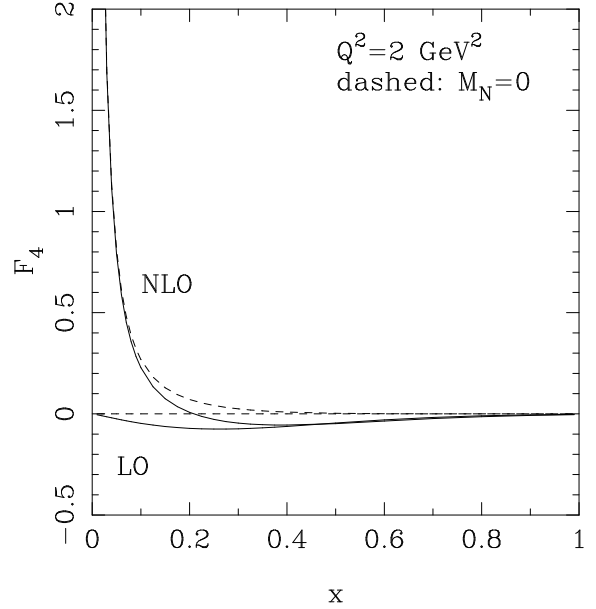}
\caption{The LO and NLO structure function $F_4$ as a function of $x$ for
$Q^2=2\ {\rm GeV}^2$. The CTEQ6
parton distribution functions are used. Dashed lines show the
case with target mass $M_N=0$.
\label{fig:f4}
}
\end{figure}
Our focus here is on the DIS contribution, however, there is an issue
of double counting if one does the inclusive sum and adds to 
separate exclusive processes important in the few GeV energy region.
One way to approximately avoid double counting \cite{Lipari:1994pz}
is to impose
a cut on the hadronic energy 
\begin{equation}
W^2 = (p_N+q)^2 > (1.4\ {\rm GeV})^2
\end{equation}
for $W$-boson momentum $q$. In addition, 
in Ref.~\cite{kretzerreno} we evaluate the contributions
from a range of $Q_{min}^2$ to assess the importance of 
non-perturbative or higher-twist
contributions to the cross section.

A general hadronic tensor has six structure functions $W_i$ which
can be rescaled into the more conventional $F_i$. The structure
function $W_6$ which multiplies the antisymmetric tensor with
$p_N$ and $q$ never appears because the tensor combination contracts
to zero with the leptonic equivalent of $W_{\mu\nu}$ in the cross
section. The full differential cross section for
neutrino CC scattering, including the lepton
mass (for definiteness set to $m_\tau$) is
\begin{eqnarray} \nonumber
\frac{d^2\sigma^{\nu(\bar{\nu})}}{dx\ dy} &=& \frac{G_F^2 M_N
E_{\nu}}{\pi(1+Q^2/M_W^2)^2}\\ \nonumber
&\times & \left\{
(y^2 x + \frac{m_{\tau}^2 y}{2 E_{\nu} M_N})
F_1^{W^\pm} \right. \\ \nonumber
&+& \left[ (1-\frac{m_{\tau}^2}{4 E_{\nu}^2})
-(1+\frac{M_N x}{2 E_{\nu}}) y\right]
F_2^{W^\pm}
\\ \nonumber
&\pm& 
\left[x y (1-\frac{y}{2})-\frac{m_{\tau}^2 y}{4 E_{\nu} M_N}\right]
F_3^{W^\pm} \\  \label{eq:nusig}
&+& 
\frac{m_{\tau}^2(m_{\tau}^2+Q^2)}{4 E_{\nu}^2 M_N^2 x} F_4^{W^\pm}
\\ \nonumber
&-& \left. \frac{m_{\tau}^2}{E_{\nu} M_N} F_5^{W^\pm}
\right\}\ .
\end{eqnarray} 
where $\{x,y,Q^2\}$ are the standard DIS kinematic variables 
related through
$Q^2 = 2 M_N E_\nu x y$ and where we have neglected factors of 
$m_\tau^2/2M_N E_\nu\cdot Q^2/M_W^2$ which come 
from the 
$q^{\mu} q^{\nu} /M_W^2$ part of the  
massive boson propagator. They 
are negligible both at low and at high neutrino energies 
so do not enter our numerics.
They
can be included by replacing:
\begin{equation}
F_i^{W^\pm} \rightarrow F_i^{W^\pm} \times \left( 1 + \epsilon_i\right)
\end{equation}
with
\begin{eqnarray} \nonumber
\epsilon_1 &=& 
\frac{m_{\tau}^2\ (Q^2+2M_W^2)}{2 M_W^4}
\\ \nonumber
\epsilon_2 &=& 
-\frac{E_{\nu}^2\ m_{\tau}^2\ y\ [ 4 M_W^2+y (Q^2+m_{\tau}^2)]}{M_W^4\ [4(y-1)E_{\nu}^2+m_{\tau}^2+Q^2]}
\\ 
\epsilon_3 &=& 0 \\ \nonumber
\epsilon_4 &=& 
\frac{Q^2\ (Q^2+2 M_W^2)}{M_W^4}
\\ \nonumber
\epsilon_5 &=&
\frac{Q^2}{M_W^2}+\frac{(M_W^2+Q^2)\ (m_{\tau}^2+Q^2)\ y}{2 M_W^4}
\end{eqnarray}

\begin{figure}[t]
\psfig{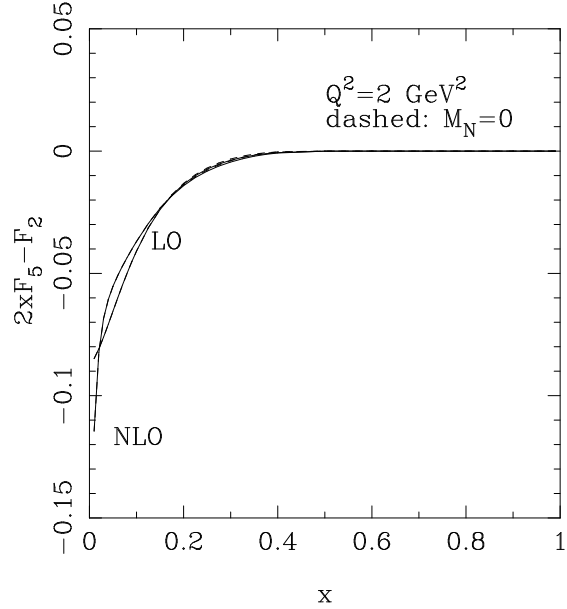}
\caption{The LO and NLO structure function difference $2xF_5-F_2$ as a function
of $x$ for
$Q^2=2$ GeV$^2$. The CTEQ6
parton distribution functions are used. Dashed lines show the
case with target mass $M_N=0$.
\label{fig:f5}
}
\end{figure}
In Eq.~\ref{eq:nusig}, the extra structure functions $F_4$ and $F_5$
appear with powers of the charged lepton mass, and so are usually
neglected. At leading order, they are related to the usual
structure functions by the Albright-Jarlskog relations \cite{aj},
generalizations of the Callan-Gross relations:
\begin{eqnarray} \label{eq:aj1}
F_4 &  = &0 \\ \label{eq:aj2}
{2x F_5} &  = & F_2 \ ,
\end{eqnarray}
These relations are violated by
kinematic target mass corrections and 
at NLO
in QCD when quark masses are retained. 
More specifically, Eq.~\ref{eq:aj1} is violated by $M_N\neq 0$ and
${\cal O}(\alpha_s )$ corrections, but not by the charm mass $m_c$
when the strange quark mass vanishes. Eq.~\ref{eq:aj2} is violated
by non-zero $M_N$ and $m_c$, but for the massless case, this
Albright-Jarlskog relation holds to any order in $\alpha_s$.

The complete lepton mass correction comes from the above
equation together with the limits of integration to get the full
cross section. Details of the lepton mass dependence of the
integration limits appear in Ref. \cite{kretzerreno}.

Target mass corrections appear in Eq.~\ref{eq:nusig}, and they
also appear implicitly in the $F_i$'s. For neutrino scattering,
it is conventional to rewrite $F_i$ in terms of ${\cal F}_i$
which are written at leading order, neglecting the target and
charm quark masses,  as
\begin{equation}
{\cal F}_i= (1-\delta_{i4})q (x,Q^2)\ 
\end{equation}
in terms of quark distribution functions $q(x,Q^2)$.

Target mass corrections appear in the collinear 
($p_q^\perp=0$) limit due to the
fact that the {\it small} nucleon light cone momentum 
$p_N^{-}$ is not a simple rescaling of 
the massless parton momentum $p_q^{-}$ when we have
$p_N^{+} = \xi p_q^{+}$ for the {\it large} components.
When one includes $M_N$ in the collinear limit
and $m_c$, where $m_c$ is replaced by
the parameter $\lambda\equiv Q^2/(Q^2+m_c^2)$,
one gets for the case of CC charm production:
\begin{eqnarray} \label{eq:mix1}
F_1^c &=& {\cal{F}}_1^c \\ \label{eq:mix2}
F_2^c &=& 2\ \frac{x}{\lambda}\ \frac{{\cal{F}}_2^c}{\rho^2} \\ \label{eq:mix3}
F_3^c &=& 2\ \frac{{\cal{F}}_3^c}{\rho}\\ \label{eq:mix4}
F_4^c &=& \frac{1}{\lambda }\ \frac{(1-\rho )^2}{2 \rho^2 }
\ {\cal{F}}_2^c + {\cal{F}}_4^c + \frac{1-\rho}{\rho}\ {\cal{F}}_5^c \\ \label{eq:mix5}
F_5^c &=& \frac{{\cal{F}}_5^c}{\rho} - \frac{(\rho -1)}{\lambda \rho^2}
\ {\cal{F}}_2^c\ .
\end{eqnarray} 
Here,
\begin{equation}
\rho^2 \equiv 1 + \left( \frac{2 M_N x}{Q}\right)^2 \ .
\end{equation}
A further target mass correction come in through the Nachtmann
variable \cite{nacht}
$\eta = 2x/(1+\rho)$ for massless quark production and
$\tilde{\eta}=\eta/\lambda$ for charm production.
For the result shown below
\begin{equation}
{\cal{F}}_i^c=(1-\delta_{i4})q(\tilde{\eta},Q^2)+{\cal O}(\alpha_s)\ .
\end{equation}
Expressions for the ${\cal O}(\alpha_s)$ corrections appear in
detail in Ref. \cite{kretzerreno}.
The light quark contributions are obviously understood to be included 
in the $\lambda \rightarrow 1$ limit.

\section{STRUCTURE FUNCTION RESULTS}
In Figs.~\ref{fig:f4}, \ref{fig:f5} we quantify the violations of the
Albright-Jarlskog relations discussed above.
The violations are concentrated at low-$x$ which is not
the most relevant kinematic region for present neutrino energies.
The naive Albright-Jarlskog relations therefore provide reasonable
approximations in the evaluation of integrated cross sections.

The target mass corrections shown in Figs. \ref{fig:f4},\ref{fig:f5}
are in the collinear limit. These corrections differ by at most
10\%  from the target mass corrected structure functions evaluated
in the operator product expansion approach \cite{gp} in the range
of $Q^2=1-4$ GeV$^2$ \cite{soon}. The differences are smaller than 10\%  where
the structure functions are largest.

\section{CROSS SECTION RESULTS}
\begin{figure}[t]
\epsfig{figure=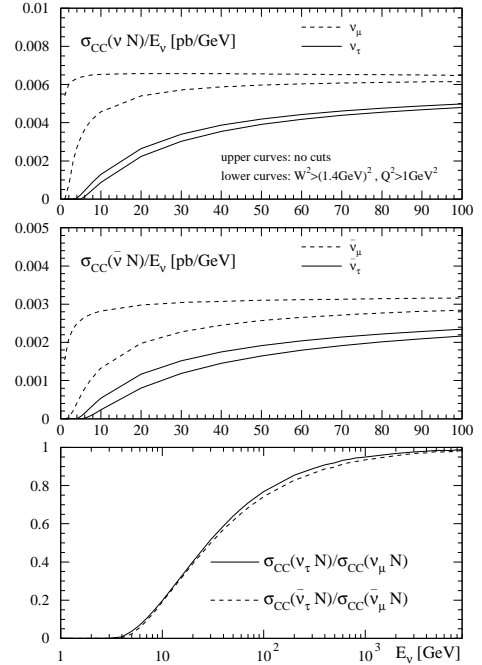,width=8cm}
\caption{Comparative results for tau 
and muon neutrino cross section versus energy with and without
various cuts applied.
\label{fig:fig6}
}
\end{figure}
The integrated cross sections for tau and muon (anti-)neutrino
DIS are plotted in Fig.~\ref{fig:fig6}. The uncertainty
in the cross sections from varying
the parton distribution functions within their bounds
is typically about a few percent: The results in 
the figures are based on CTEQ6 \cite{cteq6}. We have also compared 
to GRV98 \cite{grv98}.
\begin{figure}[t]
%fig3.eps
\epsfig{figure=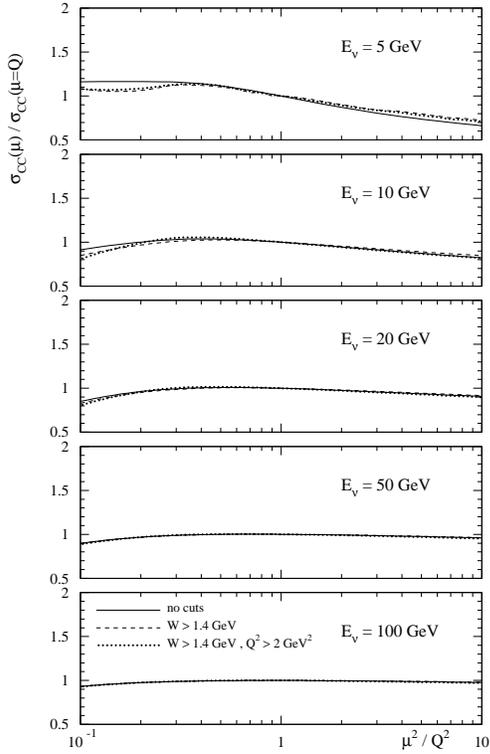,width=8cm}
\caption{The ratio 
of $\sigma_{\rm CC}(\mu)/\sigma_{\rm CC}(\mu=Q)$ in NLO for $\nu_\tau$N
interactions, as a function of $\mu^2/Q^2$ for several values of $E_\nu$.
\label{fig:mu}}
\end{figure}

\subsection{$\sigma(\nu_\tau N\rightarrow \tau X)$}
\begin{figure}[t]
\epsfig{figure=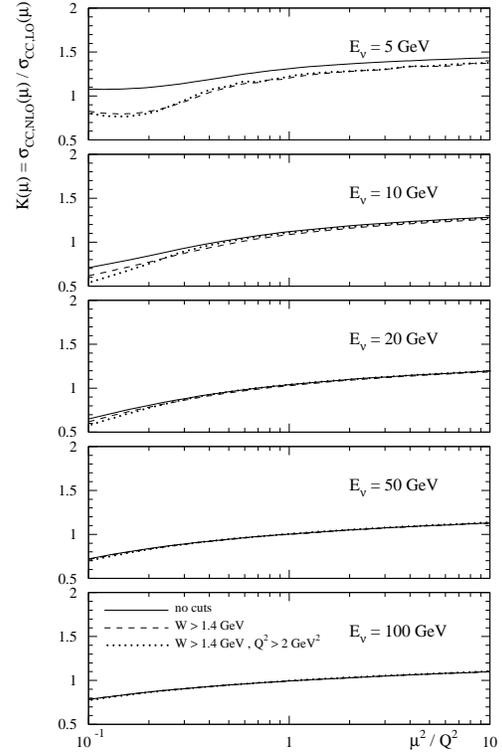,width=8cm}
%fig4.eps
\caption{The K-factor $K=$NLO/LO versus factorization 
scale $\mu$ for tau neutrinos with no cuts
(solid), $W_{\min}=1.4$ GeV (long dash) and $W_{\min}=1.4$ GeV with
$Q^2>2$ GeV$^2$ (short dash).
\label{fig:kfac2}
}
\end{figure}
In terms of a purely DIS evaluation of the neutrino cross-section,
$\tau$ production is a natural observable because the tau mass
cuts off non-DIS interaction to a good extent. The general NLO features of 
our results for
$\nu_{\tau}$ DIS are, therefore, equally valid for $\nu_{\mu}$ DIS
with kinematic DIS cuts.
In Figs.~\ref{fig:mu} and \ref{fig:kfac2} we
show two typical features of the inclusion of NLO effects:
Scale dependence and the size of the K-factor. The scale 
dependence is substantially reduced compared to a LO
calculation and the K factor hints at the importance of
NLO corrections towards low neutrino energies.

\subsection{$\sigma(\nu_\mu N\rightarrow \mu X)$}
Different from $\tau$ production, the deep inelastic component 
\begin{figure}[t]
\epsfig{figure=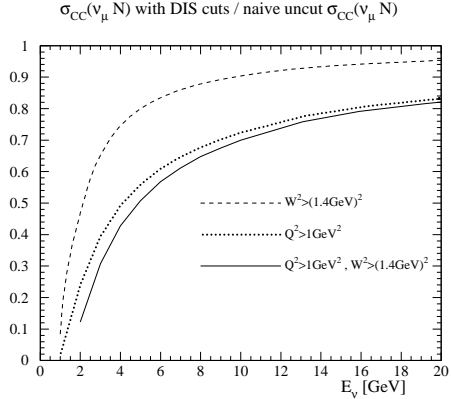,width=8cm}
%fig4.eps
\caption{The muon neutrino cross section versus energy with DIS cuts.
\label{fig:nuint1}
}
\end{figure}
becomes small in $\nu_{\mu}$ scattering at lower energies as shown in 
Fig.~\ref{fig:nuint1}. Since the tau threshold in $\nu_{\tau}$ DIS
extends to large energies one may wonder about the impact of muon mass 
terms in low energy $\nu_{\mu}$ scattering. Fig.~\ref{fig:nuint2} looks at
such corrections to the DIS component. The results strongly
depend on the kinematic cuts imposed. This observation, taken together with 
the smallness of the {\it bona fide} deep inelastic component at such energies
do not allow to draw a conclusion from the perturbative framework which we 
consider here. Rather, one will have to look into the equivalent corrections
to elastic and low-multiplicity form factors.

\section{CONCLUSIONS}
\label{sec:sum}

The $\nu N$ cross section is an important ingredient in current and future
atmospheric and neutrino factory experiments. Our evaluation 
\cite{kretzerreno}
of the
NLO corrections for $\nu_\tau$N CC interactions including charm mass
corrections and an estimate of target mass effects is part of a larger
theoretical program to understand the inelastic $\nu N$ cross section
over the full energy range relevant to experiments. 
Apart from oscillation searches, the recent observation of an anomaly
in the ratios of neutral to charged current neutrino ineractions 
also necessitate a solid theory for neutrino interactions with nuclei.
Research along these lines is progressing 
and further results are under completion \cite{soon}.
\begin{figure}[t]
\epsfig{figure=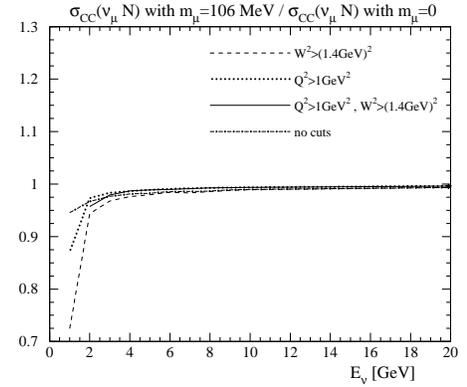,width=8cm}
%fig4.eps
\caption{The muon mass effect on the DIS neutrino cross section
with and without kinematic cuts (applied to both numerator and 
denominator).
\label{fig:nuint2}
}
\end{figure}

\end{document}